\documentclass[%
a4paper,
prd,  
twocolumn,
superscriptaddress,
nofootinbib,
nobibnotes,
amsmath,amssymb,
aps,
fleqn,
floatfix%
]{revtex4-2}
\usepackage{amsfonts,graphics,color}
\usepackage[english]{babel}
\usepackage{graphicx}
\usepackage{dcolumn}
\usepackage{bm}
\newcommand{\red}{\color{red}}
\newcommand{\redn}{\color{red}}
\newcommand{\black}{\color{black}}
\renewcommand{\red}{\black}
\renewcommand{\redn}{\black}
\usepackage[normalem]{ulem}
\definecolor{myblue}{rgb}{0.05,0.1,0.5}

\usepackage{xcolor}
\definecolor{xlinkcolor}{cmyk}{1,1,0,0}
\usepackage[bookmarks=true, pdfnewwindow=true, colorlinks=true, linkcolor=xlinkcolor, citecolor=xlinkcolor, filecolor=xlinkcolor, urlcolor=xlinkcolor, final=true]{hyperref}
 
\begin{document}
\title[Very high energy neutrinos with Baikal-GVD]{%
Constraints on the diffuse flux of multi-PeV astrophysical neutrinos obtained with the Baikal Gigaton Volume Detector 
}
\author {V.A.~Allakhverdyan}
\affiliation{ Joint Institute for Nuclear Research, Dubna, 141980  Russia}

\author{A.D.~Avrorin}
\affiliation{ Institute for Nuclear Research of the Russian Academy of Sciences, 
Moscow, 117312 Russia}

\author{A.V.~Avrorin}
\affiliation{ Institute for Nuclear Research of the Russian Academy of Sciences, 
Moscow, 117312 Russia}

\author{V.~M.~ Aynutdinov}
\affiliation{ Institute for Nuclear Research of the Russian Academy of Sciences, 
Moscow, 117312 Russia}

\author{Z.~Be\v{n}u\v{s}ov\'{a}}
\affiliation{ Comenius University, Bratislava, 81499 Slovakia}
\affiliation{ Czech Technical University, Institute of Experimental and Applied Physics, CZ-11000 Prague, Czech Republic}

\author{I.A.~Belolaptikov}
\affiliation{ Joint Institute for Nuclear Research, Dubna, 141980  Russia}

\author{E.~A.~ Bondarev}
\affiliation{ Institute for Nuclear Research of the Russian Academy of Sciences, 
Moscow, 117312 Russia}

\author{I.V.~Borina}
\affiliation{ Joint Institute for Nuclear Research, Dubna, 141980  Russia}

\author{N.M.~Budnev}
\affiliation{Irkutsk State University, Irkutsk, 664003 Russia}

\author{V.A.~Chadymov}
\affiliation{ Independed researcher}

\author{A.S.~Chepurnov}
\affiliation{ Skobeltsyn Research Institute of Nuclear Physics, Moscow State University, Moscow, 119991 Russia}

\author{V.Y.~Dik}
\affiliation{ Joint Institute for Nuclear Research, Dubna, 141980  Russia}
\affiliation{ Institute of Nuclear Physics ME RK, Almaty, 050032 Kazakhstan}

\author{A.N.~Dmitrieva}
\affiliation{ National Research Nuclear University MEPHI, Moscow, Russia, 115409}

\author{\fbox{G.V.~Domogatsky}}
\affiliation{ Institute for Nuclear Research of the Russian Academy of Sciences, 
Moscow, 117312 Russia}

\author{A.A.~Doroshenko}
\affiliation{ Institute for Nuclear Research of the Russian Academy of Sciences, 
Moscow, 117312 Russia}

\author{R.~Dvornick\'{y}}
\affiliation{ Comenius University, Bratislava, 81499 Slovakia}
\affiliation{ Czech Technical University, Institute of Experimental and Applied Physics, CZ-11000 Prague, Czech Republic}

\author{A.N.~Dyachok}
\affiliation{ Irkutsk State University, Irkutsk, 664003 Russia}

\author{Zh.-A.M.~Dzhilkibaev}%
  \affiliation{ Institute for Nuclear Research of the Russian Academy of Sciences, 
Moscow, 117312 Russia}

\author{E.~Eckerov\'{a}}
\affiliation{ Comenius University, Bratislava, 81499 Slovakia}
\affiliation{ Czech Technical University, Institute of Experimental and Applied Physics, CZ-11000 Prague, Czech Republic}

\author{T.V.~Elzhov}
\affiliation{ Joint Institute for Nuclear Research, Dubna, 141980  Russia}

\author{V.N.~Fomin}
\affiliation{ Independed researcher}

\author{A.R.~Gafarov}
\affiliation{ Irkutsk State University, Irkutsk, 664003 Russia}

\author{K.V.~Golubkov}
\affiliation{ Institute for Nuclear Research of the Russian Academy of Sciences, 
Moscow, 117312 Russia}

\author{T.I.~Gress}
\affiliation{ Irkutsk State University, Irkutsk, 664003 Russia}

\author{K.G.~Kebkal}
\affiliation{ LATENA, St. Petersburg, 199106  Russia}

\author{V.K.~Kebkal}
\affiliation{ LATENA, St. Petersburg, 199106 Russia}

\author{I.V.~Kharuk}
\affiliation{ Institute for Nuclear Research of the Russian Academy of Sciences, 
Moscow, 117312 Russia}

\author{S.S.~Khokhlov}
\affiliation{ National Research Nuclear University MEPHI, Moscow, Russia, 115409}

\author{E.V.~Khramov}
\affiliation{ Joint Institute for Nuclear Research, Dubna, 141980  Russia}

\author{M.M.~Kolbin}
\affiliation{ Joint Institute for Nuclear Research, Dubna, 141980  Russia}

\author{S.O.~Koligaev}
\affiliation{ INFRAD, Dubna, 141981  Russia}

\author{K.V.~Konischev}
\affiliation{ Joint Institute for Nuclear Research, Dubna, 141980  Russia}

\author{A.V.~Korobchenko}
\affiliation{ Joint Institute for Nuclear Research, Dubna, 141980  Russia}

\author{A.P.~Koshechkin}
\affiliation{ Institute for Nuclear Research of the Russian Academy of Sciences, 
Moscow, 117312 Russia}

\author{V.A.~Kozhin}
\affiliation{ Skobeltsyn Research Institute of Nuclear Physics, Moscow State University, Moscow, 119991 Russia}

\author{M.V.~Kruglov}
\affiliation{ Joint Institute for Nuclear Research, Dubna, 141980  Russia}

\author{V.F.~Kulepov}
\affiliation{ Nizhny Novgorod State Technical University, Nizhny Novgorod, 603950 Russia}

\author{A.A.~Kulikov}
\affiliation{ Irkutsk State University, Irkutsk, 664003 Russia}

\author{Y.E.~Lemeshev}
\affiliation{ Irkutsk State University, Irkutsk, 664003 Russia}

\author{M.V.~Lisitsin}
\affiliation{ National Research Nuclear University MEPHI, Moscow, Russia, 115409}

\author{S.V.~Lovtsov}
\affiliation{ Irkutsk State University, Irkutsk, 664003 Russia}

\author{R.R.~Mirgazov}
\affiliation{ Irkutsk State University, Irkutsk, 664003 Russia}

\author{D.V.~Naumov}
\affiliation{ Joint Institute for Nuclear Research, Dubna, 141980  Russia}

\author{A.S.~Nikolaev}
\affiliation{ Skobeltsyn Research Institute of Nuclear Physics, Moscow State University, Moscow, 119991 Russia}

\author{I.A.~Perevalova}
\affiliation{ Irkutsk State University, Irkutsk, 664003 Russia}

\author{A.A.~Petrukhin}
\affiliation{ National Research Nuclear University MEPHI, Moscow, Russia, 115409}

\author{D.P.~Petukhov}
\affiliation{ Institute for Nuclear Research of the Russian Academy of Sciences, 
Moscow, 117312 Russia}

\author{E.N.~Pliskovsky}
\affiliation{ Joint Institute for Nuclear Research, Dubna, 141980  Russia}

\author{M.I.~Rozanov}
\affiliation{ St. Petersburg State Marine Technical University, St. Petersburg, 190008 Russia}

\author{E.V.~Ryabov}
\affiliation{ Irkutsk State University, Irkutsk, 664003 Russia}

\author{G.B.~Safronov}
\affiliation{ Institute for Nuclear Research of the Russian Academy of Sciences, 
Moscow, 117312 Russia}

\author{B.A.~Shaybonov}
\affiliation{ Joint Institute for Nuclear Research, Dubna, 141980  Russia}

\author{V.Y.~Shishkin}
\affiliation{Skobeltsyn Institute of Nuclear Physics MSU, Moscow, Russia, 119991}

\author{E.V.~Shirokov}
\affiliation{ Skobeltsyn Research Institute of Nuclear Physics, Moscow State University, Moscow, 119991 Russia}

\author{F.~\v{S}imkovic}
\affiliation{ Comenius University, Bratislava, 81499 Slovakia}
\affiliation{ Czech Technical University, Institute of Experimental and Applied Physics, CZ-11000 Prague, Czech Republic}

\author{A.E. Sirenko}
\affiliation{ Joint Institute for Nuclear Research, Dubna, 141980  Russia}

\author{A.V.~Skurikhin}
\affiliation{ Skobeltsyn Research Institute of Nuclear Physics, Moscow State University, Moscow, 119991 Russia}

\author{A.G.~Solovjev}
\affiliation{ Joint Institute for Nuclear Research, Dubna, 141980  Russia}

\author{M.N.~Sorokovikov}
\affiliation{ Joint Institute for Nuclear Research, Dubna, 141980  Russia}

\author{I.~\v{S}tekl}
\affiliation{ Czech Technical University, Institute of Experimental and Applied Physics, CZ-11000 Prague, Czech Republic}

\author{A.P.~Stromakov}
\affiliation{ Institute for Nuclear Research of the Russian Academy of Sciences, 
Moscow, 117312 Russia}

\author{O.V.~Suvorova}
   \affiliation{ Institute for Nuclear Research of the Russian Academy of Sciences, 
Moscow, 117312 Russia}

\author{V.A.~Tabolenko}
\affiliation{ Irkutsk State University, Irkutsk, 664003 Russia}

\author{V.I.~Tretjak}
\affiliation{ Joint Institute for Nuclear Research, Dubna, 141980  Russia}

\author{B.B.~Ulzutuev}
\affiliation{ Joint Institute for Nuclear Research, Dubna, 141980  Russia}

\author{Y.V.~Yablokova}
\affiliation{ Joint Institute for Nuclear Research, Dubna, 141980  Russia}

\author{D.N.~Zaborov}
\affiliation{ Institute for Nuclear Research of the Russian Academy of Sciences, 
Moscow, 117312 Russia} 

\author{S.I.~Zavjalov}
\affiliation{ Joint Institute for Nuclear Research, Dubna, 141980  Russia}

\author{D.Y.~Zvezdov}
\affiliation{ Joint Institute for Nuclear Research, Dubna, 141980  Russia}

\collaboration{Baikal--GVD Collaboration}

\author{A.~V.~Plavin}
\affiliation{ Black Hole Initiative at Harvard University, 
Cambridge, MA 02138, USA}

\author{D.~V.~Semikoz}
\affiliation{ APC, Universit\'e Paris Diderot, CNRS/IN2P3, CEA/IRFU, Sorbonne Paris Cit\'e, 119 75205 Paris, France}

\author{S.~V.~Troitsky}
\thanks{Corresponding author, e-mail: st@inr.ac.ru}
\affiliation{ Institute for Nuclear Research of the Russian Academy of Sciences, 
Moscow, 117312 Russia}
\affiliation{ Physics Department, Lomonosov Moscow State University, 
Moscow 119991, Russia}
\date{Submitted to \textit{Physical Review D} on July 9, 2025}%

\begin{abstract}
Various theoretical models predict cosmic neutrinos with multi-PeV energies. The recent detection of a $\sim 10^{17}$~eV neutrino in the KM3NeT experiment suggests that these energetic particles can be studied with present-day installations. Here, we present upper limits on the flux of astrophysical neutrinos with energies $(10^{15.5} - 10^{20})$~eV obtained with the largest liquid-water neutrino telescope, Baikal Gigaton Volume Detector (GVD), operation using cascade-like events. We discuss astrophysical implications of these results and constrain several cosmogenic neutrino scenarios using a combination of Baikal-GVD, KM3NeT, IceCube and Auger data.
\end{abstract}

\maketitle


\section{Introduction}
\label{sec:Introduction}
Ultra-high energy (UHE) cosmic rays (CR) are accelerated in astrophysical sources and propagate towards the Earth from cosmological distances. Interactions of UHE particles with the Cosmic Microwave Background (CMB) radiation become important at high energies due to pion production for protons and spallation for heavier nuclei. For the uniform distribution of UHECR sources in the Universe, these interactions result in the suppression of the observed cosmic-ray flux above $\sim 10^{19.5}$~eV, known as the Greizen-Zatsepin-Kuzmin (GZK) cutoff \cite{Greisen:1966jv,Zatsepin:1966jv}. Such a suppression in the UHECR spectrum was observed by the High Resolution Fly's Eye (HiRes) experiment \cite{HiRes:cutoff}, and later confirmed by the Pierre Auger Observatory (Auger) \cite{PierreAuger:cutoff} and Telescope Array (TA) \cite{TelescopeArray:cutoff}. While the shape of the cutoff observed by HiRes and TA is consistent with the GZK prediction for the proton-dominated primary composition, the Auger observations favor the ``disappointing scenario'' \cite{disappointing} in which the suppression is caused by the maximal energy achieved in cosmic acceleration, and the composition is heavier. 

These very same interactions produce secondary gamma rays and neutrinos from $\pi$-meson decays. These cosmogenic neutrinos, pioneered by Berezinsky and Zatsepin \cite{BerezinskyZatsepin:cosmogenic}, are now considered as an important diagnostic tool for UHECR models, see e.g.\ Ref.~\cite{Aloisio:cosmogenic}. Indeed, the determination of the energy spectrum and the primary composition of highest-energy cosmic rays is plagued by systematic uncertainties related to the lack of firm understanding of hadronic interactions at energies beyond the reach of collider experiments, cf.\ e.g.\ Refs.~\cite{TA-1.27spectrum,Auger-hadronic,WHISP-2023}. The cosmogenic neutrino flux, if observed, could be used to infer cosmic-ray composition and source parameters in an independent way. Roughly, higher fluxes of cosmogenic neutrinos are associated with lighter composition and/or excess of sources at high redshifts, but the actual fluxes depend on many parameters, see e.g.\ Refs.~\cite{Kalashev:neutrino2002,Ehlert:cosmogenic} and references therein. Cosmogenic neutrinos are expected to come from all UHECRs in the Universe, providing truly diffuse flux of neutrinos, as opposite to neutrinos from individual astrophysical sources.

\red
However, this cosmogenic neutrino flux is yet to be measured. Only three neutrino events with best-fit reconstructed energies above 5~PeV have been detected by the presently largest neutrino telescope, IceCube. The highest energy of these three is about 13~PeV, see e.g.\ Ref.~\cite{Muzio:above5PeV} for details and further references.
\black
Recently, the KM3Net experiment has reported about the detection of a  neutrino with best-fit energy well above 100~PeV \cite{KM3NeT:Nature}. This event could potentially be the first cosmogenic neutrino ever detected \cite{KM3NeT:cosmogenic}. However, to make conclusions from this observation, one should take into account that events of similar energies have not been detected by other neutrino telescopes, despite larger exposure. These non-detections constrain strongly the isotropic neutrino flux, see e.g.\ \cite{KM3NeT:agnostic,titans}. 

Still under construction, Baikal-GVD \cite{Baikal-JETP} is the largest operating neutrino telescope in the Northern hemisphere, and has already collected significant exposure for high-energy neutrinos. No event with energy above $10^{15.5}$~eV has been detected by Baikal-GVD in the cascade mode so far, so we present here corresponding upper limits on the isotropic flux of neutrinos at highest energies, and discuss possible implications of this non-detection.

\section{Data}
\label{sec:data}
Baikal-GVD is a water Cerenkov neutrino telescope, currently having the instrumented volume $\sim 0.7$~km$^3$ and growing. It is located in Lake Baikal, Russia, at the latitude of $51.8^\circ$ North. 

\begin{figure}
\centering
\includegraphics[width=0.95\linewidth]{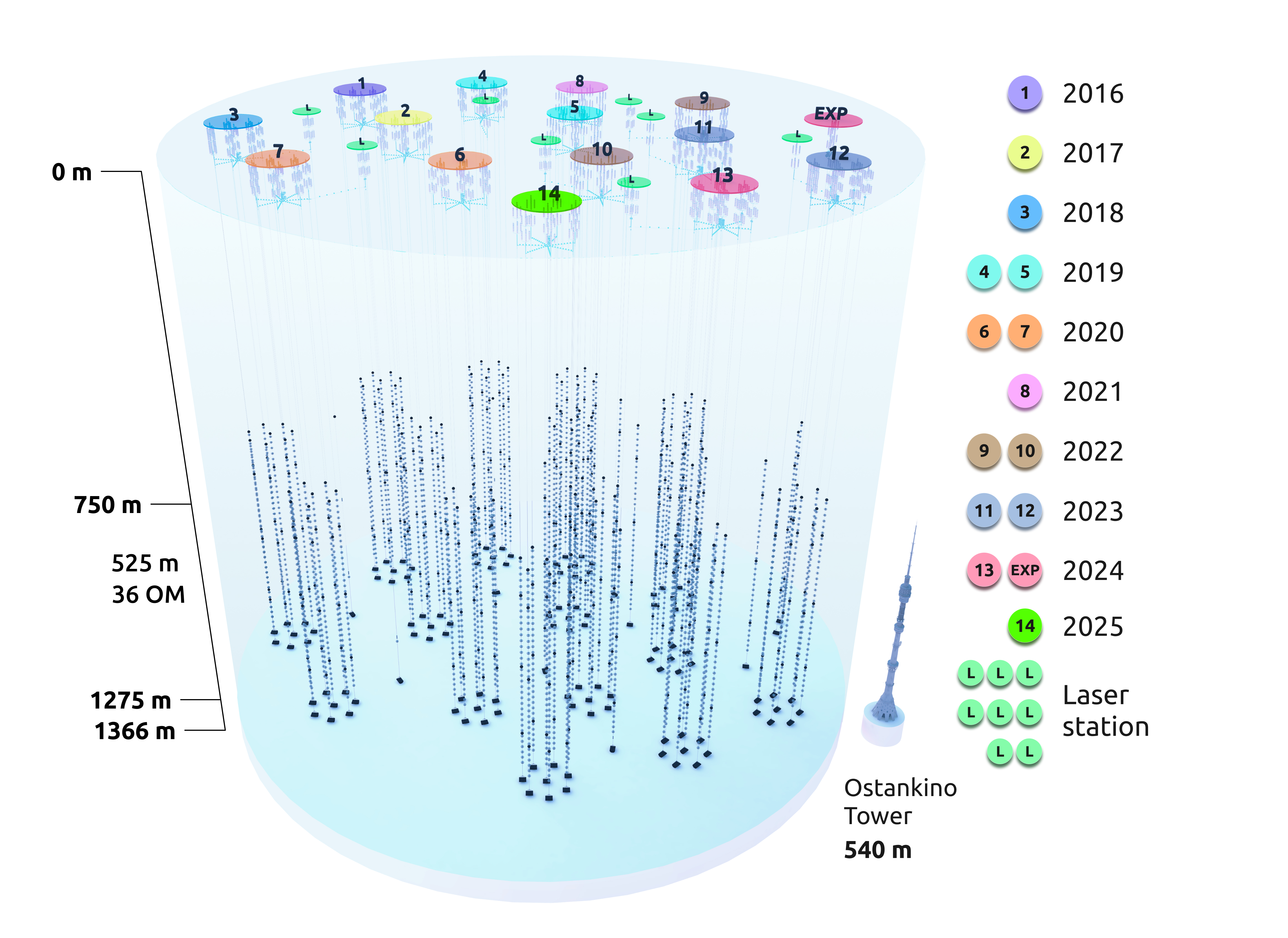}
\caption{\label{fig:detector}
\red Schematic view of the Baikal-GVD configuration achieved by April 2025. The present work uses data from clusters 1--12.
}
\end{figure}

For the present analysis, we use the Baikal-GVD sample of cascade events detected between Spring 2018 and Spring 2024. In this period, the detector was operated in several, increasingly larger configurations, with one or two new ``clusters'' of optical modules (OMs) added each spring. One cluster comprises 288 OMs arranged on eight vertical strings, with 36 OMs per string. \red The layout of the detector is presented in Fig.~\ref{fig:detector}.
The inter-cluster spacing is varied in the range 250--300 m. Given the typical geometrical size of the cascade, the installation effectively works as a set of independent clusters in the cascade mode.   
\black
Over the stated period, the number of clusters increased from three to \red twelve\black. The total exposure of the data set used here corresponds to $\approx 26.8$ years of one-cluster operation. The reader may find descriptions of the experiment, event reconstruction, analysis techniques and procedures e.g.\ in Refs.~\cite{Baikal-JETP,Baikal-diffuse,Baikal-tracks,Baikal-TXS,Baikal-diffuse2025}.

\red
The search strategy for high-energy neutrino detection with Baikal-GVD was described in Ref.~\cite{Baikal-diffuse}. High-energy cascade events with the multiplicity of triggered optical modules (OMs) $N_{\rm hit}>$19 and reconstructed energy $E_{\rm sh} > 70$~TeV were selected, and additional cuts which suppress events from atmospheric muons were applied. The cut $N_{\rm hit}>$19 was optimised for effective suppression of misreconstructed background events from atmospheric muon bundles. The number of these background events decrees rapidly with energy due to the steep energy spectrum of atmospheric muons. 

In the present work, we select high-energy cascade events with $N_{\rm hit}>$8  and impose the quality cuts described in Ref.~\cite{Baikal-diffuse}, except for the cut introduced to suppress atmospheric muon bundles. Relaxing this latter cut and multiplicity of OMs allows us to increase the exposure considerably. Simulations indicate that the number of background events is about $1.3$ for all energies above 3~PeV, and $<0.1$ above 10~PeV with these relaxed cuts.
\black

\begin{figure}
\centering
\includegraphics[width=0.95\linewidth]{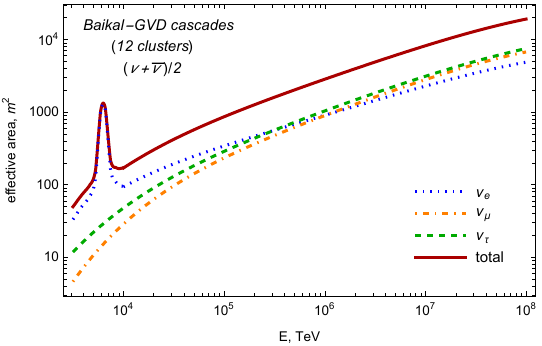}
\caption{\label{fig:effarea}
Effective area of Baikal-GVD for the detection of very high energy neutrino-induced cascades as a function of the neutrino energy (exposure-weighted average over the upper hemisphere\red, $2\pi$ solid angle\black).
}
\end{figure}

The effective area of Baikal-GVD as of 2023, determined by Monte-Carlo simulations following Ref.~\cite{Baikal-diffuse}, is presented in Fig.~\ref{fig:effarea}. 
Note that for cascade events produced by electron (anti)neutrinos, the Landau-Pomeranchuk-Migdal (LPM) effect suppresses the effective area at high energies. The LPM effect was taken into account in simulations following Refs.~\cite{LPM97,LPM98,LPM99}. For water, it starts to suppress bremsstrahlung for electron energies above $\sim 2$~PeV and therefore significantly reduces the effective area for electron (anti)neutrinos with energies above $\sim 100$~PeV. The peak of the effective area about 6.3~PeV is related to the resonant production of $W$ bosons by electron antineutrinos, known as the Glashow resonance.

\section{Flux limits}
\label{sec:limits}
\begin{figure*}
\centering
\includegraphics[width=0.8\linewidth]{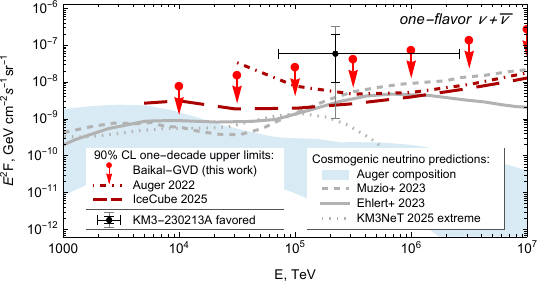}
\caption{\label{fig:spec}
Very high energy diffuse neutrino fluxes per one flavor, sum of neutrinos and antineutrinos (assuming isotropic flux and flavor equipartition). Red arrows: upper limits from Baikal-GVD cascades (this work). Dash-dotted dark red line: limits from Pierre Auger Observatory \cite{Auger:limit2022}. Long-dashed dark red line: limits from IceCube \cite{IceCube:2025ezc}. All limits are 90\% CL and obtained in decade-wide bins \red assuming $1/E$ spectrum in the bin \black (the Auger limit was \red recalculated \black to match th\red{ese} \black  convention\red{s}\black). Black point with error bars: value favored by the observation of KM3-230213A event \cite{KM3NeT:Nature} (black -- 90\% CL in energy, 68\% CL in flux; gray -- 95\% CL in flux). For comparison, some of the predictions of the cosmogenic neutrino flux are shown; gray band -- baseline predictions based on Auger spectrum and composition model \cite{Auger:cosmogenic2023}, short-dashed gray line -- maximal predicted flux from Ref.~\cite{Muzio:cosmogenic}, full gray line -- best-fit flux in the ``2SC-dip'' model of Ref.~\cite{Ehlert:cosmogenic}, dotted gray line -- the extreme source evolution model of Ref.~\cite{KM3NeT:cosmogenic}.
}
\end{figure*}

The highest-energy event found in the Baikal-GVD cascade sample had the energy of $\approx 1200$~TeV; this event has already been reported in Ref.~\cite{Baikal-diffuse}. There are no events in the range of interest, $E>10^{3.5}$~TeV. The expected atmospheric background is negligible, so we use the standard approach for the Poisson statistics, recommended by Ref.~\cite{PDG}, to obtain 90\% CL upper limit on the Poisson mean of the observed number of events as $N<N_{90}\approx2.3$. As it is customary in high-energy neutrino astrophysics, we derive our limit for overlapping decade-wide energy bins, assuming $1/E$ spectrum within each bin to account for the energy dependence of the exposure, that is
\begin{equation}
\int_{E_{\rm min}}^{E_{\rm max}} \! dE \int\! dt \,\red d \Omega \black \, A(E,t\red,\Omega\black) \, F < N_{90},
\label{eq:integral}
\end{equation}
where $F \propto 1/E$, $A(E,t\red,\Omega\black)$ is the effective area, the integration extends over the bins with $E_{\rm max}=10E_{\rm min}$ in energy\red, over the celestial sphere in the solid angle $\Omega$ \black and over the data collection time in $t$. \red The effective area, measured in m$^2$, is estimated by Monte-Carlo simulations and takes into account the efficiency of neutrino detection for a particular set of selection cuts. \black
The resulting upper limits are listed in Table~\ref{tab:bins}
\begin{table}
\begin{center}
\begin{tabular}{cc}
\hline\hline
energy bin,  &~ $E^2F$ upper limit, \\
TeV &~ $10^{-8}$\,GeV\,cm$^{-2}$s$^{-1}$sr$^{-1}$\\
\hline
$10^{3.5} - 10^{4.5}$ & 0.78 \\
$10^{4.0} - 10^{5.0}$ & 1.6 \\
$10^{4.5} - 10^{5.5}$ & 2.5 \\
$10^{5.0} - 10^{6.0}$ & 4.3 \\
$10^{5.5} - 10^{6.5}$ & 7.6 \\
$10^{6.0} - 10^{7.0}$ & 14 \\
$10^{6.5} - 10^{7.5}$ & 27 \\
$10^{7.0} - 10^{8.0}$ & 54 \\
\hline\hline
\end{tabular}
\end{center}
\caption{\label{tab:bins} Baikal-GVD 90\% CL upper limits on the diffuse astrophysical neutrino flux per one flavor, sum of neutrinos and antineutrinos (assuming isotropic flux and flavor equipartition). }
\end{table}
and presented graphically in Fig.~\ref{fig:spec}.
\section{Discussion}
\label{sec:disc}
\subsection{Baikal-GVD limits and results of other experiments}
\label{sec:disc:comparison}
Baikal-GVD upper limits on the isotropic neutrino flux are compared in Fig.~\ref{fig:spec} with the observation by KM3NeT \cite{KM3NeT:Nature} and latest upper limits from IceCube \cite{IceCube:2025ezc} and Auger \cite{Auger:limit2022}. Baikal-GVD limits are becoming stronger as compared to Auger limits at the energies below 100~PeV. Our upper limit is comparable to that of IceCube in the lowest energy bin, and is not in tension with the IceCube observation of several events at those energies \cite{IceCube:2025ezc}.     

\begin{figure}
\centering
\includegraphics[width=0.95\linewidth]{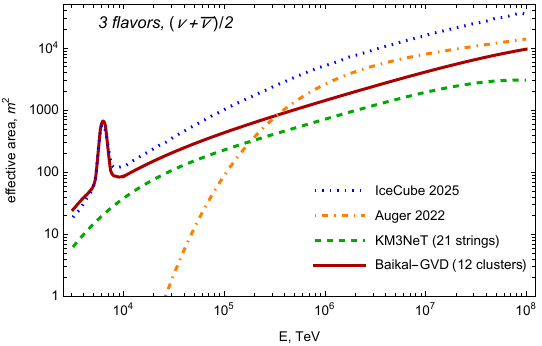}
\caption{\label{fig:compare-effarea}
Effective areas of Baikal-GVD (cascades, 2023 configuration, full red line, this work), KM3NeT (ARCA bright track selection, 2023 configuration, Ref.~\cite{KM3NeT:Nature}, dashed green line), Auger (2022 analysis in Ref.~\cite{Auger:limit2022}, compiled in Ref.~\cite{KM3NeT:agnostic} based on the data release accompanying Ref.~\cite{Auger:effarea}, dot-dashed orange line), and IceCube (2025 analysis, Ref.~\cite{IceCube:2025ezc}, dotted blue line) for the detection of very high energy neutrinos (exposure-weighted full-sky\red, $4\pi$ solid angle, \black average, three flavors, average over neutrino and antineutrino).
}
\end{figure}

\begin{figure*}
\centering
\includegraphics[width=0.8\linewidth]{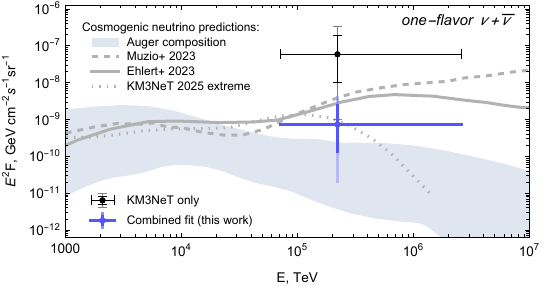}
\caption{\label{fig:specfit}
Very high energy diffuse neutrino fluxes per one flavor, sum of neutrinos and antineutrinos (assuming isotropic flux and flavor equipartition). The blue cross represents the result of the combined fit of four experiments (darker -- 90\% CL in energy, 68\% CL in flux; lighter -- 95\% CL in flux), see the text for details. Other notations are the same as in Figure~\ref{fig:spec}.  
}
\end{figure*}

Following Ref.~\cite{KM3NeT:agnostic}, we estimate the neutrino flux at the energies between 72~PeV and 2600~PeV, that is in the 90\%~CL energy interval of KM3-230213A. We assume that this event originates from the isotropic, steady $E^{-2}$ flux of neutrinos equally distributed between flavors. The energy-dependent effective areas for KM3NeT \cite{KM3NeT:Nature} (335 days exposure), IceCube \cite{IceCube:2025ezc} (4605 days), Auger \cite{Auger:limit2022} (18 years), and Baikal-GVD (this work; 26.8 cluster-years) are presented in Fig.~\ref{fig:compare-effarea}.
Integrating the total exposure over energy, assuming $E^{-2}$ neutrino flux, and having one observed event for zero background, we obtain the estimate
$$
F^{(\nu+\bar\nu)}_{\rm 1-flavor} = \left(7.2^{+16.5}_{-5.9} \right) \times 10^{-10} ~ \rm{GeV}\,{\rm cm}^{-2} {\rm s}^{-1} {\rm sr}^{-1}.
$$
It may be compared with the KM3NeT-only estimate in Fig.~\ref{fig:specfit}.
\red
This estimate is slightly lower than that of Ref.~\cite{KM3NeT:agnostic}, $7.5 \times 10^{-10}~ \rm{GeV}\,{\rm cm}^{-2} {\rm s}^{-1} {\rm sr}^{-1}$, because of the extra exposure brought into the analysis (for other post-KM3NeT works on combination of very high energy neutrino fluxes, see e.g.\ Refs.~\cite{titans,Muzio:above5PeV}). \black
\subsection{Constraints on the cosmogenic neutrino flux models}
\label{sec:disc:cosmogenic}
Predictions of the cosmogenic neutrino flux depend strongly on the assumed properties (injection spectra and composition) and spatial distribution (cosmological evolution and number density) of UHECR sources, as well as on the chosen model of the cosmic photon background, see e.g.\ \cite{Kalashev:neutrino2002}. From a plethora of available theoretical predictions, we have chosen five recent models to compare with the new experimental constraints.

\paragraph{Auger 2023 low and high.} Ref.~\cite{Auger:cosmogenic2023} presents a number of cosmogenic neutrino fluxes derived from simultaneous fits to spectra and composition observed by Auger, assuming two components from different populations of sources with varying parameters. These predictions span a wide band between the lowest and highest fluxes, both corresponding to the strong positive evolution, $(1+z)^5$, of the low-energy component with the redshift $z$, and the flat evolution of the high-energy component, but differing by assumed maximal source redshifts, $z_{\rm max}=1$ for the low-flux model versus $z_{\rm max}=5$ for the high-flux one.

\paragraph{Muzio et al.\ 2023 maximal.}
Ref.~\cite{Muzio:cosmogenic} considers a particular model in which cosmic rays interact efficiently near their sources, and assume a significant injected proton component at super-GZK energies. We use the model providing for the maximal neutrino flux, which corresponds to the AGN evolution and EPOS-LHC interaction model.

\paragraph{Ehlert et al.\ 2023 2SC-dip.}
Ref.~\cite{Ehlert:cosmogenic} performs a comprehensive study of possibilities to have a significant UHECR proton component without contradicting to observational data. Interestingly, their UHECR best fit within one of the two models considered, 2SC-dip, is in tension with observational constraints on the neutrino flux. We use this best-fit model prediction without multmessenger constraints because they are being reconsidered in the present study.

\paragraph{KM3NeT 2025 extreme evolution. }
Ref.~\cite{KM3NeT:cosmogenic} extended the source-evolution parameter space in the Auger data fit to obtain the extreme scenario with strong positive evolution and $z_{\rm max}=6$, in which cosmogenic neutrinos could explain the KM3-230213A event, which we also consider.

Predictions of the selected models are shown in gray color in both Fig.~\ref{fig:spec} and Fig.~\ref{fig:specfit}. To obtain quantitative constraints on the scenarios, we \red make use of Eq.~(\ref{eq:integral}) and substitute there the theoretically predicted flux $F$. In this way, we calculate, for each of the models, the expected number of events $N_{\rm th}$ in the KM3-230213A 90\% CL energy uncertainty band, in which the observed number of events is one. These theoretical numbers are presented in Table~\ref{tab:cosmogenic-norms}. \black
\begin{table}
\begin{center}
\begin{tabular}{cccc}
\hline\hline
model  &\red $N_{\rm th}$&$k$ & $k_{\rm f}$\\
\hline
Auger low 2023 \cite{Auger:cosmogenic2023}&\red 0.006&$<$610& $160^{+360}_{-130}$  \\
Auger high 2023 \cite{Auger:cosmogenic2023}&\red 0.34&$<$11.5& $3.0^{+7.0}_{-2.4}$ \\
Muzio$+$ maximal 2023 \cite{Muzio:cosmogenic}&\red 7.0& $<$0.56& $0.14^{+0.33}_{-0.12}$ \\
Ehlert$+$ dip 2023 \cite{Ehlert:cosmogenic}&\red 4.1& $<$0.96& $0.25^{+0.60}_{-0.20}$ \\
KM3NeT 2025 extreme \cite{KM3NeT:cosmogenic}&\red 0.98& $<$3.9& $1.0^{+2.3}_{-0.8}$ \\
\hline\hline
\end{tabular}
\end{center}
\caption{\label{tab:cosmogenic-norms} \red 
Model cosmogenic neutrino fluxes constrained from the combined fit of Auger, Baikal-GVD, IceCube and KM3NeT. \red $N_{\rm th}$ is the expected number of events between 72~PeV and 2600~PeV for the model flux with normalization $k=1$. Other columns give the 90\% CL upper limit on $k$ and, under assumption of the cosmogenic origin of the observed event, its 68\% confidence interval $k_{\rm f}$.}
\end{table}
\red
Assuming that the model cosmogenic neutrino spectrum scales with the normalization factor $k$, where $k=1$ for the theoretical model and the spectral shape does not change, we obtain 90\% CL upper limits on $k$ from Eq.~(\ref{eq:integral}). We then make a temporary assumption that the single observed event was cosmogenic, and use the Poisson statistics \cite{PDG} to get the best-fit value of $k$ and its 68\% confidence interval. These numbers are also presented in Table~\ref{tab:cosmogenic-norms}. We see that the assumption of the cosmogenic origin of KM3-230213A is in tension with the lowest-flux Auger 2023 model, with the Poisson p-value corresponding to \redn about $2.5\sigma$ \black discrepancy.
\black

\section{Conclusions}
\label{sec:concl}
This work presents upper limits on the isotropic astrophysical neutrino flux at energies above $10^{3.5}$~TeV from searches of neutrino-induced cascades in the 2018--2023 data of Baikal-GVD in the construction stage, equivalent to $\sim 2$ years of the present configuration of the detector. In the lower part of the energy band, these limits are of the same order as those reported by IceCube, and supersede those published by Auger. We further concentrate on the energy range in which the KM3-230213A event was detected. We combine the exposure of Baikal-GVD with those of IceCube, Auger, and KM3NeT, and present the best-fit isotropic $E^{-2}$ flux estimate from the joint analysis. 
With the combined exposure, we also obtain constraints on selected models of cosmogenic neutrinos; other models for the isotropic  neutrino flux can be tested in the same way with the help of the presented energy-dependent effective areas. Future analyses are needed to understand the origin of KM3-230213A and to obtain more information on UHECR sources with the neutrino data.

\section*{Acknowledgements}
This work is supported in the framework of the State project ``Science'' by the Ministry of Science and Higher Education of the Russian Federation under the contract 075-15-2024-541.
The work of A.V.P.\ was supported by the Black Hole Initiative, which is funded by grants from the John Templeton Foundation (Grant \#60477, 61479, 62286) and the Gordon and Betty Moore Foundation (Grant GBMF-8273).

\section*{Data availability}
The effective area of the Baikal-GVD experiment is available in an ancillary file with this paper's arXiv version \cite{ThisPaper-arXiv}.
\bibliography{GVD-VHE}
\end{document}